\documentclass[manuscript,screen]{acmart}

\AtBeginDocument{%
  }

\usepackage{multirow}
\usepackage[most]{tcolorbox}
\usepackage{xcolor}
\usepackage{threeparttable}
\usepackage[most]{tcolorbox}
\usepackage{xcolor}

\newtcbox{\functag}{on line,
  enhanced,
  rounded corners,
  boxrule=0.5pt,
  colframe=black!45,
  colback=black!8,
  left=5pt,right=5pt,top=2pt,bottom=2pt,
  boxsep=0pt,
  before upper=\strut,
  fontupper=\bfseries\scriptsize
}

\newtcbox{\stacktag}{on line,
  enhanced,
  rounded corners,
  boxrule=0.5pt,
  colframe=black!35,
  colback=black!4,
  left=5pt,right=5pt,top=2pt,bottom=2pt,
  boxsep=0pt,
  before upper=\strut,
  fontupper=\scriptsize
}

\tcbuselibrary{skins}

\newtcolorbox{rqbox}{
  enhanced,
  colback=white,
  colframe=black!20,
  rounded corners,
  boxrule=0.6pt,
  left=3pt,right=3pt,top=3pt,bottom=3pt,
  colbacklower=black!5,
  fontlower=\small,
  middle=3pt,
  segmentation style={draw=none},
  before skip=6pt, after skip=6pt
}

\newtcolorbox{scenariobox}{
  colback=black!5,
  colframe=black!5,
  sharp corners,
  boxrule=0pt,
  left=6pt,right=6pt,top=4pt,bottom=4pt,
  borderline west={2pt}{0pt}{black!25},
  fontupper=\small
}
\renewcommand\footnotetextcopyrightpermission[1]{}
\renewcommand\footnotetextcopyrightpermission[1]{}

\begin{document}

\title{Operationalizing NCII and CSAM Law in the Age of AI: Open Problems}

\author{Kimberly Mai}
\affiliation{%
  \institution{AI Security Institute}
  \country{UK}
}

\author{Cate Heine}
\affiliation{%
  \institution{AI Security Institute}
  \country{UK}}
\email{cate.heine@dsit.gov.uk}
\correspondingauthor

\author{Vinith M. Suriyakumar}
\affiliation{%
  \institution{AI Security Institute}
  \country{UK}
}
\affiliation{%
  \institution{MIT}
  \country{USA}
}

\author{Ashia C. Wilson}
\affiliation{%
 \institution{MIT}
 \country{USA}}

\renewcommand{\shortauthors}{Mai et al.}

\begin{abstract}
Advances in generative AI have driven a surge in synthetic non-consensual intimate imagery (NCII) and child sexual abuse material (CSAM), prompting a series of legal reforms. These reforms rely on legal definitions where technical operationalization is ambiguous, such as whether a depicted person is identifiable, whether a model is capable of or intended for abuse generation, and whether an abusive image can be traced to the model that produced it. Drawing on the law of England and Wales and the United States, we examine the relevant law alongside the technical research needed to operationalize it. From this, we develop a research agenda of open problems at this technical-legal interface, spanning the interpretation of existing law, the prevention of harm before an AI tool is deployed, and enforcement of offenses. These are problems the technical community is well placed to solve, and where progress is vital to real-world protection.
\end{abstract}

\authorsaddresses{%
  Corresponding author: Cate Heine, AI Security Institute, UK;
  \href{mailto:cate.heine@dsit.gov.uk}{cate.heine@dsit.gov.uk}.%
}
\maketitle

\makeatletter
\renewcommand{\footrulewidth}{0pt}
\fancyfoot{}                       
\makeatother
\thispagestyle{empty}

\section{Introduction}
AI-generated sexual abuse imagery is increasing sharply, targeting both children and adults. The Internet Watch Foundation reported a 26,385\% increase in AI-generated child sexual abuse videos between 2024 and 2025 \cite{iwf2026aicsam}, while another report found that websites allowing users to nudify images of real people receive over 40 million unique visitors per month \cite{isd2026mapping}. These harms are real and enduring. Generative models trained and finetuned on abuse imagery revictimize survivors and normalize extreme violence \cite{iwf2026aicsam}.

Approaches to tackle AI-generated non-consensual intimate imagery (NCII) and child sexual abuse material (CSAM) are inherently sociotechnical. Defenses range from technical guardrails that block explicit requests to generate this imagery, to laws that criminalize its creation and distribution. Lawmakers across the world have moved to expand existing laws and create new ones to cover synthetic imagery \cite{icmec2023csam, parti2024legal}. However, these reforms rely on legal definitions where technical operationalization is ambiguous. For example, new laws ban the creation of AI models specialized for generating NCII or CSAM, but determining whether a model is specialized for abuse generation is an open technical question.

Our aim is not to comment on these legal choices, but to examine what technical developments would help to operationalize them. Such research is currently sparse. Differences in the legal definitions of AI-generated NCII and CSAM across jurisdictions, legal barriers to safeguards testing, and impacts on the welfare of researchers have made it an underexplored area of study compared to other areas within digital trust and safety.  We therefore set out a research agenda to guide this work. Our contributions are as follows:

\begin{enumerate}
    \item We analyze NCII and CSAM laws in England and Wales and the United States, and their applicability to AI-generated imagery.
    \item We identify and taxonomize open technical questions that emerge where legal definitions do not clearly translate to existing technical operationalizations. 
    \item We map the taxonomy to the deployment stack to aid in practical implementation.
\end{enumerate}

\section{Background}

\subsection{Definitions}

This paper covers two types of sexual abuse imagery, divided by the perceived age of the depicted person: NCII and CSAM. NCII and CSAM cover both real and AI-generated content. In this paper, we focus on images and videos generated by, or with the assistance of, an AI system. While these categories are legally and conceptually distinct, they share commonalities in the technical mechanisms used to produce the material and hence raise overlapping technical problems.

NCII is intimate imagery of an adult that is produced or shared without that person’s consent. It requires the person to be real and identifiable. CSAM is material depicting the sexual abuse of a child. It encompasses a broad range of material, though this paper covers images and videos. We use the most widely accepted international definition that a child is a person under eighteen years old \cite{icmec2023csam}. Unlike adult intimate imagery, CSAM is never consensual. Additionally, some definitions of CSAM do not require a real child: a fully synthetic depiction can constitute CSAM.

We also distinguish between the terms `tool' and `service' in this paper, as they raise different operationalization challenges. A tool is something a user runs and configures directly, e.g. open-source software a user installs on their own hardware. A service is something a third party manages and the user accesses through an interface or API, such as an image-editing app available through an app store. We use `model' to refer to the trained artifact (weights and architecture) that generates images or video, distinct from the tool or service through which a user accesses it.

\subsection{Policy and Legal Landscape}
\label{subsec:policylandscape}

Exact definitions of CSAM differ across jurisdictions \cite{icmec2023csam}. The legal status of NCII is more nascent, with some laws already in place and others in active development \cite{isd2026ncii}. As this paper focuses more on identifying technical operationalization gaps rather than comparing laws, we ground our legal analysis in two jurisdictions: England and Wales and the United States.

Across these jurisdictions, criminal law may address several types of acts relating to NCII and CSAM. We identify five:
(1) creating content, (2) possessing content, (3) distributing content, (4) creating or providing a tool or service used to produce content, and (5) creating or providing materials that train others on how to produce content.

Laws criminalize these acts to varying degrees, sometimes combining them into a single offense, sometimes addressing them separately or not covering them at all. Beyond these five, we also consider whether a jurisdiction offers a testing exemption: legal protections for parties to conduct good faith research on NCII and CSAM \cite{longpre2024position}. For each category of NCII and CSAM, we examine coverage across both jurisdictions. Table \ref{tab:legal-overview} provides a summary.

\begin{table}[t]
  \caption{Overview of legal coverage in England and Wales (E\&W), and the U.S.}
  \label{tab:legal-overview}
  \small
  \begin{tabular}{@{}lcccccc@{}}
    \toprule
    & \multicolumn{3}{c}{CSAM} & \multicolumn{3}{c}{NCII} \\
    \cmidrule(lr){2-4} \cmidrule(l){5-7}
    Provision & E\&W & U.S. fed. & U.S. state & E\&W & U.S. fed. & U.S. state \\
    \midrule
    Creating content            & Yes & Yes                & Yes                & Yes & No  & Partial\textsuperscript{b} \\
    Possessing content          & Yes & Partial\textsuperscript{a} & Partial\textsuperscript{b} & No & No  & Partial\textsuperscript{b} \\
    Distributing content        & Yes & Yes                & Yes                & Yes & Yes & Yes \\
    Creating tools or services  & Yes & No                 & Partial\textsuperscript{b} & Yes & No  & Partial\textsuperscript{b} \\
    Training materials          & Yes & No                 & No                 & No  & No  & No \\
    Testing exemption           & In progress\textsuperscript{c} & No & No & In progress\textsuperscript{c} & No & No \\
    \bottomrule
  \end{tabular}

  \vspace{2pt}
  \raggedright\footnotesize
  \textsuperscript{a} Covers AI-generated material where it depicts an identifiable real child, is indistinguishable from a real child, or is obscene.
  \textsuperscript{b} Covered in some states but not others; scope and definitions vary.
  \textsuperscript{c} Commenced under Section 111 of the Crime and Policing Act 2026 but not yet operational.
\end{table}

\subsubsection{England and Wales}
England and Wales (E\&W) constitutes a single criminal jurisdiction. It has legislated across all five types of acts, as well as including a testing exemption. Therefore, we discuss these developments first before turning to the U.S. landscape.

\textbf{CSAM}. The creation, possession, and distribution of CSAM is illegal. Distinctions are made by how photorealistic the content is. Section 1 of the Protection of Children Act 1978 criminalizes the taking, distribution, and possession of an `indecent photograph or pseudo photograph of a child'. The specific term `pseudo photograph' is the definition that extends the Act to photorealistic AI-generated content. Section 62 of the Coroners and Justice Act 2009 criminalizes `prohibited images' of children, including non-realistic AI-generated images, cartoons, illustrations, and animations. The definitions of `indecent' versus `prohibited' cover different thresholds. Indecent imagery covers the photorealistic depiction of erotic poses or sexual imagery, while prohibited imagery differs in the content needed to meet the threshold. Beyond these legal definitions, the Internet Watch Foundation and law enforcement categorize images by severity and nature \cite{iwf2026aicsam}.

The possession of training materials for creating CSAM is also an offense under Section 69 of the Serious Crime Act 2015, which extends to materials assisting the creation of AI-generated imagery. The Crime and Policing Act 2026 introduced a new measure to criminalize the creation, possession, and distribution of `CSA image generators’. This measure covers programs and services that are produced with the intention of creating CSAM, rather than those that are capable of doing so. 

The same 2026 Act also introduces a `technology testing defence’ under Section 111, which allows the government to authorize approved organizations to test an AI’s system's capability to produce CSAM. This defence is currently being operationalized.

\textbf{NCII}. Criminal offenses around NCII are more recent. The sharing or threat of sharing intimate imagery where there is reasonable belief that the depicted person does not consent is an offense under Sections 66A-66D of the Sexual Offences Act 2003, as amended by the Online Safety Act 2023. The creation of intimate images without consent was addressed more recently through the Data (Use and Access) Act 2025. The law is written in a way that it is neutral to whether the image is real or synthetic.

Section 66I of the Sexual Offences Act 2003, as amended by the Crime and Policing Act 2026 also criminalizes NCII image generators (referred to as a `purported intimate image generator'), but there is no equivalent offense for creating training materials to produce NCII. The technology testing defence includes NCII in scope.

\textbf{Regulation}. Alongside offenses targeting individuals, there are also regulations that place duties on platforms and services that may host or facilitate CSAM and NCII. The primary regulation is the Online Safety Act 2023, which requires user-to-user and search services to take proportionate measures to prevent users from encountering priority illegal content. Once the entities are aware of priority illegal content, they must take `reasonable steps to take the content down'. CSAM and NCII are designated priority offenses under the Act and hence attract the most stringent duties.

The designated regulator, Ofcom, publishes guidance operationalizing these definitions for services. For example, Ofcom’s definition of a photograph or film includes `data which can be converted to a photograph or film' as equivalent, and includes URLs linking to NCII and CSAM in scope \cite{ofcom2026guidance}, extending obligations beyond hosting images directly.

\subsubsection{United States}
\label{sec:us-law}
U.S. law addresses NCII and CSAM through a combination of federal statute and state laws. Coverage is narrower than in E\&W, particularly for models and training materials.

\textbf{CSAM}. The creation and distribution of synthetic CSAM is illegal at the federal level and in most states \cite{enoughabuse}. In \textit{Ashcroft v. Free Speech Coalition} 535 U.S. 234 (2002), the Supreme Court held that purely synthetic imagery is protected speech unless it is obscene or depicts an actual, identifiable child. In response, Congress passed the PROTECT Act 2003, clarifying the prohibition on synthetic CSAM to obscene depictions under the \textit{Miller v. California} 413 U.S. 15 (1973) test. The test establishes three criteria for obscenity: whether an average person applying contemporary community standards would find the material appeals to prurient interest, whether the material is patently offensive, and whether the work lacks serious literary, artistic, political, or scientific value. These criteria point to a contextual judgment rather than a property that can be read off an image. 

Federal CSAM law covers images specifically, not auxiliary acts such as the creation of training materials. Some states, however, have started to criminalize tools or services used to produce CSAM. The Texas Responsible AI Governance Act 2025 prohibits developing or distributing an AI system with the sole intent of producing CSAM; similarly, Tennessee's HB 0769 criminalizes knowingly possessing, distributing, or producing tools and services designed for the purpose of creating CSAM. Minnesota's HF 1606 takes a different approach, prohibiting a person who owns or controls a website, software, or other service from allowing users to nudify an image of an identifiable individual (including minors), doing so on the user's behalf, or advertising such services. The prohibition does not apply where the service requires the technical skill of a user to nudify, defined as the `substantial application of individualized technological or artistic skill by a human creator'. In this sense, Texas focuses on intent and Tennessee on design, both mirroring E\&W's intent-based framing, while Minnesota focuses on the function the service performs rather than its intended purpose.

There are currently no testing exemptions that permit approved organizations to test whether an AI system can produce CSAM. %check

\textbf{NCII}. At the federal level, the TAKE IT DOWN Act 2025 criminalizes the distribution of NCII, but not creation or possession. The Act also extends to children, but only where the depicted individual is identifiable. Treatment of NCII otherwise varies at the state level, with some states amending existing laws and others introducing new legislation.

As with CSAM, there are no offenses covering training materials, though the Texas Responsible AI Governance Act 2025 and Minnesota’s HF 1606 extend to NCII as well as CSAM. 

The U.S. does not have a statutory testing mechanism or testing exemption comparable to the E\&W technology testing defence.

\textbf{Regulation}. The U.S. has two mechanisms at the regulatory level. Firstly, under 18 U.S.C. § 2258A (2018), platforms must report to NCMEC's CyberTipline upon becoming aware of CSAM. Secondly, the TAKE IT DOWN Act 2025 created a new obligation for platforms to act on valid removal requests for NCII and CSAM within 48 hours and to make reasonable efforts to identify known identical copies. The Federal Trade Commission, which enforces TAKE IT DOWN, maintains guidance on this duty.

\subsection{Relevant Synthetic Generation Techniques}
A range of AI techniques could facilitate the creation of NCII and CSAM. None of these are inherently malicious, but each could be misused.

\textbf{Generation techniques} include face swaps, inpainting, style transfer and complete synthetic generation. The first three rely on an existing image of a real person, whereas complete synthetic generation may not involve a real individual at all, meaning these generation techniques map unevenly onto existing laws and mitigations. We describe the techniques in more detail in Appendix \ref{appendix:generation_techniques}. We use the term `nudification' to refer to the use of generation or editing techniques to remove or alter clothing from otherwise benign imagery of a person.

\textbf{Customization techniques}.
The generation techniques above can be used off-the-shelf or customized. Finetuning adjusts a model by training it on examples of the desired output. A popular finetuning approach is low-rank adaptation (LoRA) \cite{hu2022lora}. LoRAs are add-on files that update a limited set of weights rather than the full model. They can be trained using a few images on consumer hardware and shared independently of the base model \cite{hawkins2025deepfakes}. Although they are used for legitimate purposes, LoRAs can adapt a general purpose model so it specializes in generating NCII and CSAM. They often circulate as standalone files, with one study identifying 35,000 variants capable of generating non-consensual images of individuals on public model hosting platforms \cite{hawkins2025deepfakes, wei2024exploring}. 

\section{Open Technical Problems}
\label{sec:technical}
We derive the open questions of this section from the legal analysis in §\ref{subsec:policylandscape}. For each area, we identify where the application of the legal requirement depends on a technical judgment that current methods do not clearly provide. We organize these questions around the five components addressed in §\ref{subsec:policylandscape} as well as a section on ethics that could support research more broadly. For each component, we propose a set of relevant technical questions that would help with operationalizing the law. 

We tag each question across two further dimensions. Firstly, we tag the question by the function its answer serves:
\begin{itemize}
    \item \textbf{Interpretation}. Clarifying how an existing law could be applied.
    \item \textbf{Prevention}. Enabling interventions to stop generation or distribution before harm occurs.
    \item \textbf{Enforcement}. Enabling detection, attribution, or evidencing of a specific offense.
\end{itemize}

Secondly, we tag each question by the relevant part of the deployment stack: training data, base model, customization, inference, distribution, or detection \cite{isd2026mapping, gibson2025analyzing}. A question may span more than one part.

Each open question covers a relevant operational scenario, prior research, and directions for potential research. The work most similar to ours is \cite{openproblemschildsafety2026}, which sets out a research agenda on open problems for generative AI child safety. Our agenda complements theirs and differs in how the problems are derived and scope: we ground each question in specific legal constructs across two jurisdictions and identify the technical problems that arise in applying them. We also broaden the scope to include NCII, which raises distinct questions around identifiability, and identify additional problems beyond those they raise.

\subsection{Creation of Content}
\label{subsec:question_creation}

\begin{rqbox}
\subsubsection{What are potential ways to monitor the usage and activity around open weight models used to create NCII or CSAM?}
\label{subsubsec:monitor}
\par
\begin{flushleft}
\functag{Enforcement}
\stacktag{Distribution}
\stacktag{Detection}
\end{flushleft}
\tcblower
   \textbf{Scenario:} A regulator wants to track how widely an open weight model is being used to produce NCII and CSAM to prioritize interventions, but once downloaded, the model's usage is harder to track.
\end{rqbox}

Closed weight service providers can monitor usage at the point of access. They can detect malicious prompts and block illegal outputs through web apps or APIs. In contrast, open weight developers lose visibility on models once downloaded. As a result, understanding usage depends on indirect signals: where a model is hosted, who downloads it, how it is modified and distributed, and what its outputs look like once shared. Two sources of signals are the ecosystem where models circulate and the outputs a model produces.

At the ecosystem level, some partial signals of these activities exist on the public web. Model hosting platforms record download counts, model relationships (e.g. declared base models), and user-uploaded example images, which prior work has used to characterize malicious models and LoRAs on model marketplaces \cite{hawkins2025deepfakes,wei2024exploring}. Ecosystem studies have mapped the infrastructure behind nudification services, including the hosting, payment, and distribution channels they rely on \cite{ding2026stop,gibson2025analyzing,isd2026mapping} and how techniques and targets shift over time \cite{cui2026celebrities}. However, many of these signals are self-reported, easily falsified, and disappear when models move to file-sharing sites, private forums, or messaging channels. Open weight developers often require users to share their contact information and agree to acceptable use policies before download; these steps can record who obtains a model but cannot enforce constraints on what is done with it after download.

At the output level, tracing an image to the model that produced it can be difficult, as those sharing AI-generated images rarely disclose how they made them. To aid in this, some model providers include watermarks in generated images---small, often subtle or invisible signals which indicate that the image was produced by a particular model. 

Existing research is mixed as to whether watermarks can be removed by finetuning \cite{hu2024stable,gao2025toward,wang2025sleepermark,qi2026cert}. When present, developers can use watermarks to monitor the distribution of malicious content created with their models but cannot directly investigate the private creation of content. Law enforcement, on the other hand, may be able to infer private creation of content from a given watermark alongside other contextual clues: for example, a watermarked image found on a seized device shows which model produced it, which can help separate creation from possession of downloaded images.

Open problems remain at both levels---which ecosystem signals are predictive of malicious use, what output-level signals such as watermarks, fingerprints or metadata can reliably attribute an image to a model and survive evasion (\ref{subsubsec:fingerprint}), and how to design monitoring that adapts to continually shifting ecosystems to avoid detection.

\subsection{Possession of Content}
\label{subsec:question_possession}

\begin{rqbox}
\subsubsection{How can we detect if an image contains a real, identifiable person?}
\par
\begin{flushleft}
\functag{Interpretation}
\stacktag{Inference}
\stacktag{Detection}
\end{flushleft}
\tcblower
\textbf{Scenario:} An adversary has been found to be in possession of an AI-generated intimate image in a jurisdiction where possessing NCII is illegal. It is unclear if the image's subject is fictional, which would not classify as NCII, or a real person, which would classify as NCII.
\end{rqbox}

NCII requires a depiction of an individual who has not consented, meaning they must be identifiable. Identifiability is not just important for evidence, but also for determining whether an offense has occurred. However, existing guidance does not specify how similar a generated output must be to a target to be identifiable, what features can be used for identification, or if automated identification tools need to make similar decisions to a human investigator.

The degree of similarity between an image and a person is contested and depends on the generation technique. For face swaps and inpainting, the output may include a person’s real likeness by construction. For fully synthetic generation, this boundary is more uncertain, as a person's face may be replicated or only partially replicated in generated images. \cite{shahreza2024unveiling} studied six synthetic face recognition systems and found that synthetic faces can be mapped to real faces in the training data across all datasets using a similarity threshold measure. However, they argue that a similarity threshold is insufficient proof of leakage. On qualitative human review, they found false positives such as matches with no visible faces and matches where reviewers disagreed that the faces were similar. Forensics research shows that attributes beyond faces could be used for identification, such as skin marks, knuckle and vein patterns, and tattoos \cite{nurhudatiana2013individuality, hrkac2016tattoo, dan2023persistence}. However, it is unclear if these features have been studied in synthetic imagery and whether they remain reliable as identifiers. Finally, automated tools may not identify people the way that humans do. \cite{kang2026if} found that human judgments of identity diverge from computational similarity metrics, although they created the stimuli using non-deep learning techniques.

Existing research on these criteria does not focus on NCII specifically. Future work clarifying how to measure identifiability, particularly in cases where facial features are obscured or not present, would help with operationalizing a detection tool.

\begin{rqbox}
\subsubsection{Are model weights themselves convertible to images?}
\label{subsubsec:convertible}
\par
\begin{flushleft}
\functag{Interpretation}
\stacktag{Base model}
\stacktag{Customization}
\stacktag{Inference}
\end{flushleft}
\tcblower
    \textbf{Scenario:} Law enforcement has found model weights on a hard drive that consistently output CSAM in a jurisdiction where the possession of images is illegal but the possession of generators is not covered explicitly by law. Do the weights fall directly under CSAM laws covering images in the same way that a URL or a compressed image might?
\end{rqbox}

Guidance in both jurisdictions treats data that can be converted into images as equivalent to the images themselves \cite{ofcom2026guidance, doj2026citizensguide}. For example, the U.S. DoJ’s guidance states that `data that can be converted into a visual image' counts as CSAM \cite{doj2026citizensguide}. This clarification prevents adversaries from using formats such as URLs or zip files to evade possession offenses. These definitions also affect the range of regulatory and enforcement actions available. If a model’s weights count as data convertible to images, existing image offenses could apply to models directly. \cite{gorwa2024moderating} discusses whether hosting a model is equivalent to hosting content, and whether content moderation actions should apply to models. However, the conversion principle is not straightforward for models. For URLs and zip files, the stored data maps one-to-one onto the image it encodes. In contrast, a model’s weights encode a learned function over the training data distribution. Therefore, the technical question is whether a specific image can be recovered from model weights, which would point to convertibility.

Some evidence supporting this comes from work on compression and memorization. \cite{franceschelli2024training, deletang2024language} frame model training as compression, showing that models are competitive with standard algorithms on compression tasks. These results suggest that training data is encoded in the weights. However, these papers focus on language models rather than image generators, and compression does not imply recoverability. Memorization, the ability to reconstruct a near-exact copy of the training example, would make recovery possible \cite{cooper2025files}. However, whether memorization is sufficient to establish recoverability for NCII and CSAM offenses is unclear. Research into memorization has focused on other domains such as privacy and copyright \cite{hamburg2024llms, cooper2025files}. Policy developments in these domains suggest that recoverability is only plausible if it is easy to do \cite{hamburg2024llms}. \cite{cooper2025extracting} show that memorization depends on the model and the type of content, while \cite{carlini2023extracting} only found 109 near copies among 350,000 of the most-duplicated training examples, suggesting recoverable memorization is uncommon and concentrated in duplicated data.

Much of this research has focused on LLMs and base diffusion models rather than LoRAs. Research showing that LoRAs can be classified by their specialization suggests that weights carry semantic information \cite{africa2026detecting, suriyakumar2026evaluation}, but this is weaker than showing images themselves can be recovered. Systematically characterizing whether imagery can be reliably recovered from models and extending studies to LoRAs would clarify if existing offenses can extend to models or if new measures are required.

\subsection{Distribution of Content}
\label{subsec:question_distribution}

\begin{rqbox}
\subsubsection{How can we detect and take down models specialized to create NCII of an identifiable victim?}
\label{subsubsec:detecting_takedown_ncii}
\par
\begin{flushleft}
\functag{Enforcement}
\stacktag{Inference}
\stacktag{Detection}
\end{flushleft}
\tcblower
\textbf{Scenario:} Based on the name and description of a model hosted on a model hosting platform, a possible victim suspects that the model has been trained to generate NCII of them. How does the platform verify their request to remove the model?
\end{rqbox}

Platforms can already identify and take down instances of NCII across social media using perceptual hashes, compact `fingerprints' that match visually similar images \cite{stopncii}. However, there are no equivalent mechanisms for detecting or taking down models specialized to create NCII of an identifiable victim, which would support the enforcement of NCII generator offenses and regulatory action. Existing methods for detecting CSAM-specialized models could plausibly be modified, but they detect specialization to a category rather than an individual (§\ref{subsubsec:capability_detection}).

This gives rise to two related but distinct questions. Firstly, can one flag a model specialized for NCII when the victim is unknown? Metadata or extracted prompts could provide some insight, but adversaries may not always use obvious prompts for customization \cite{qiu2026discovering}. Secondly, is it possible to detect specialization given information about a specific victim? The most straightforward approach would be sampling from a model and comparing the outputs to a reference image \cite{deng2019arcface}. However, sampling from a suspected NCII model risks creating new intimate imagery of the victim. One alternative would be to operate on the weights directly (§\ref{subsubsec:convertible}), and some research has found that training on images of an identity can be partially reconstructed from LoRA weights alone \cite{yao2024risks}. This result suggests a reference image could be matched against weights directly, for instance by projecting it into a learned weight space \cite{dravid2024interpreting}.

Detection is only one part of the problem. Even where a model or service is flagged, the mechanisms for take down are not necessarily clear, particularly as models move across platforms \cite{cuevas2026deepfake}. Developing verification and takedown mechanisms that operate on models that also offer a route for victims to initiate requests without requiring them to relive their experiences is also an open problem \cite{chen2022trauma}.

\begin{rqbox}
\subsubsection{How can we reliably fingerprint and trace models despite modifications?}
\label{subsubsec:fingerprint}
\par
\begin{flushleft}
\functag{Enforcement}
\stacktag{Distribution}
\stacktag{Detection}
\end{flushleft}
\tcblower
\textbf{Scenario:} Intelligence suggests there is a set of finetuned models in circulation that are used to produce CSAM. Law enforcement wants to determine if the set of models share a common origin.
\end{rqbox}

Model weights are large and unwieldy to analyze directly. The ability to fingerprint a model (to obtain a more compact representation that distinguishes it and persists through modification) has many operational purposes, two of which we cover. The first use is around model heritage: linking a derivative model to one it came from, to understand what modifications were made. If there is a problematic parent model, removing it would raise the barrier to producing further malicious generators. Secondly, fingerprints could support with matching---recognizing a malicious model even after it has been altered.

There is a growing body of work that studies model fingerprinting for neural network classifiers, though their reliability has not been studied at scale. Several methods have been proposed to identify fingerprints of the decision boundary of deep neural network classifiers \cite{cao2021ipguard, chen2022copy, jia2021zest}, which have been used to determine whether one model was derived from a specific parent \cite{chen2022teacher}. These techniques have since been extended to generative models. Initial works used adversarial examples to establish provenance for LLMs and diffusion models \cite{gubri2024trap, jin2024proflingo, guo2024one}. However, this line of technique has been shown to be unreliable and easily circumvented for both types of architectures \cite{yang2026challenge, ha2026identifying}. More recent works have moved towards analyzing both the output distributions and hidden layer representation distributions of the original and derivative models \cite{ha2026identifying, zhang2024promptla}. 

These methods remain nascent, and more research is needed to build reliable methods that withstand a variety of potential modifications to bypass these techniques. It is unclear how easily output signals can be obfuscated, and how different modifications---incidental or deliberate attempts at evasion---degrade a fingerprint.

\subsection{Tools and Services Used to Produce Content}
\label{subsec:question_tools_services}

\begin{rqbox}
\subsubsection{Is it possible to reliably determine if a model has been trained on real NCII or CSAM?}
\label{subsubsec:data_detection}
\par
\begin{flushleft}
\functag{Enforcement}
\stacktag{Training data}
\end{flushleft}
\tcblower
\textbf{Scenario:} Law enforcement has identified a suspected CSAM generator model on an adversary's hard drive. In order for this to be considered illegal, they must find evidence that it was specialized specifically for the purpose of generating CSAM. Identifying known CSAM in the training data would be one avenue. Is this possible?
\end{rqbox}

CSAM laws in the U.S. and the definition of NCII in both jurisdictions hinge on whether images include real or identifiable individuals. However, applying this same standard to generators---which would involve determining whether imagery of real or identifiable individuals was included in training data---is an open technical problem.

There is a large and growing literature on methods for testing whether a given model has been trained on a specific piece of data, such as a specific image. These methods are called membership inference attacks (MIAs) \cite{ran2025lora}. Previous work has pointed out that use of MIAs in a legal or regulatory enforcement setting would require high confidence at very low false positive rates \cite{carlini2023extracting}, which existing MIA methods do not consistently deliver \cite{duan2023diffusion,dubinski2024towards}. Results are more promising for the specific use case of identifying model adapters that have been finetuned on NCII and CSAM: studies find MIAs are more successful on small, overfit finetunes \cite{dubinski2024towards,luo2025privacy,ran2025lora}, and NCII and CSAM LoRAs often fall into this category. However, further work is required to improve the reliability and robustness of MIAs. 

When assessing a model suspected of being trained on illegal images, it may be more useful to ask whether it was trained more broadly on images of a specific individual rather than on a specific image, as the latter is more restrictive. \cite{vora2025identity} extend the MIA literature by introducing a method for \textit{identity inference}---a direction which is directly relevant to NCII and CSAM, and could be explored further. Specifically, enforcing laws surrounding illegal generators may require showing not just that a model was trained on a specific identity, but that it was trained on that identity in combination with nude or sexualized imagery, which is currently an open technical problem.

Both membership and identity inference require knowing which image or individual to look for in the training data, which may not be available when assessing a suspected malicious model. `Extraction' methods for reconstructing memorized training data could reveal known CSAM in training data (§\ref{subsubsec:convertible}), but methods currently require known prompts \cite{carlini2023extracting,wu2024leveraging} or substantial knowledge of training data \cite{yao2024risks}. Extending this work to enable reliable reconstruction of training data in settings where knowledge is more limited would be valuable in detecting whether models have been trained on real NCII and CSAM.

\begin{rqbox}
\subsubsection{Can we classify whether a model is specialized for creating NCII and CSAM without generating outputs?}
\label{subsubsec:capability_detection}
\par
\begin{flushleft}
\functag{Enforcement}
\stacktag{Inference}
\stacktag{Detection}
\end{flushleft}
\tcblower
\textbf{Scenario:} A model hosting platform wishes to prevent NCII and CSAM generators from being uploaded to their site. In order to do so, they want to screen models upon upload in a way that is scalable and reliable but does not risk producing illegal content.
\end{rqbox}

Detecting whether a model is specialized for NCII and CSAM without generating output would enable more proactive moderation, particularly on LoRAs and finetuned model variants. Model hosting platforms could scan models, LoRAs, and demonstrations proactively before making them public. It would also support the welfare of trust and safety specialists and investigators, who could triage models without generating harmful content, and provide evidence relevant to whether a model was specialized to produce illegal material.

Prior work has shown it is possible to detect if a model is specialized for harmful tasks without generating output. \cite{suriyakumar2026evaluation} introduce the Gaussian probing approach to detect LoRAs specialized for CSAM. The method samples Gaussian noise as probes, feeds them through the diffusion model and LoRA to extract activations, and classifies the LoRAs based on the activations. \cite{yuan2025lurks} introduce a loss-based approach. They collect images representing concepts a model is not expected to generate, calculate the reconstruction losses from the model of interest, and use these losses to create an outlier detection model. 

The methods outlined above look at harmful concepts or categories, not identities. A more challenging subproblem is detecting specialization to a specific individual, which is key to identifying NCII (§\ref{subsubsec:detecting_takedown_ncii}). Research suggests that identity information may be recoverable from diffusion models. One approach incorporates identity as an explicit architectural object that can be injected into the image generation process \cite{li2024photomaker, wang2024instantid}. Other work suggests diffusion models contain interpretable semantic latent spaces \cite{kwon2023diffusion}, and semantics are usually found in the cross-attention layers \cite{hertz2023prompttoprompt, tang2023daam}, which could extend to identity information. However, in practice, it is often unclear what methods the tool's creator has used and if they rely on these mechanisms.

Future work could look at extending beyond LoRAs to other model types and ensuring robustness to deliberate evasion. For detecting specialization for NCII, future work could connect the identity-embedding work to the non-generative detection methods. For example, studying whether activation or loss-based approaches carry identifiable information, whether the confidence attached to such representations is suitable for a moderation tool, and if so, where in the architecture this information resides.

\begin{rqbox}
\subsubsection{How should research on capability detection distinguish between image editing operations and general image generation to prevent nudification?}
\par
\begin{flushleft}
\functag{Interpretation}
\stacktag{Inference}
\end{flushleft}
\tcblower
\textbf{Scenario: }An app store wants to prevent nudification services from being offered on their platform. How should they go about distinguishing nudification services from general-purpose image editing apps?
\end{rqbox}

Enforcing bans on nudification technology (such as Minnesota's HF1606) requires isolating nudification as a capability, but the operations it relies on are individually benign and commonplace in legitimate tools and services. Furthermore, previous work has found that operations such as inpainting may induce capabilities such as nudity previously thought to have been successfully unlearned, indicating that individual assessment of capabilities may not accurately reflect capabilities of combined operations \cite{lu2026concepts}. 

The most straightforward way to test for nudification capability is direct elicitation: for example, \cite{daffalla2026dual} directly queried 155 face swap services for face swaps into nude images. Of these services, 70\% complied with nudification requests, and 30\% had some kind of guardrail leading to refusal. SafeGen-Bench formalizes this approach into a benchmark, directly testing text-to-video models with combinations of text and image prompts which are individually benign but still likely to produce malicious outputs when combined \cite{ma2026safegen}. While they do not focus on nudification specifically, three of its ten categories---celebrity, privacy, and sexual content---are directly relevant.

However, direct approaches risk producing illegal imagery. Ecosystem-level monitoring (§\ref{subsubsec:monitor}) avoids this but does not substitute capability detection. It relies on tool or platform developers openly expressing that they can or should be used for nudification, while many tools used for creation of nude imagery are not labeled in this way \cite{wei2024exploring}. General purpose image editing services used directly for nudification would also fall outside this scope. Frameworks for isolating nudification capabilities are an open research area.

\begin{rqbox}
\subsubsection{How can we systematically measure the combination of capabilities that can create NCII and CSAM?}
\label{subsubsec:combination}
\par
\begin{flushleft}
\functag{Prevention}
\stacktag{Inference}
\end{flushleft}
\tcblower
\textbf{Scenario: }Investigators seize a device containing several models or tools. No single component is a definitive NCII or CSAM generator, but used together they can produce it. They need to determine which components contributed to the capability.
\end{rqbox}

Offenses criminalizing generators may assume that the ability to produce NCII and CSAM is self-contained in a single artifact. However, an adversary may construct this capability by combining components that are benign in isolation. For example, an adversary may combine a general inpainting model with a LoRA specialized to an individual. In this example, no single component is an NCII and CSAM generator. Identifying which components contribute most strongly to the harmful capability would help with digital forensics, helping investigators reconstruct the tools an adversary combined, and would help target mitigations more precisely.

\cite{okawa2023compositional} propose a theoretical framework for studying how compositional capabilities emerge. They find compositional capability is driven by the multiplicative influence of individual capabilities and emerges once these individual capabilities are reliably learned. Benchmarks like ConceptMix and ConceptMix++ measure real image generation \cite{wu2024conceptmix, gan2025conceptmix++}, finding output quality degrades as the number of concepts increases, and is sensitive to prompt variation---indicating prompts are not a reliable indicator of output-level capabilities.

Research on compositionality so far has not focused on such material and examines capabilities on the base models directly. It is unclear how to measure the contributions of different capabilities if they are distributed across a base model and add-ons such as LoRAs, and how to vary capabilities systematically when prompts are an unreliable way of doing so.

\begin{rqbox}
\subsubsection{Can intent to produce NCII or CSAM be inferred from a model directly?}
\label{rq:intent-vs-capability}
\par
\begin{flushleft}
\functag{Interpretation}
\stacktag{Training data}
\stacktag{Base model}
\stacktag{Customization}
\stacktag{Inference}
\end{flushleft}
\tcblower
\textbf{Scenario: }Investigators seize a device containing an image generator but with little surrounding context---no prompts or output files. To bring a prosecution, they must establish that the model was intended for producing NCII or CSAM, beyond capability.
\end{rqbox}

Laws criminalizing NCII and CSAM generators define a generator by the intent behind it (e.g. E\&W’s Crime and Policing Act and Texas’s Responsible AI Governance Act). However, it is not always straightforward to say whether a generator is or is not intended for NCII and CSAM. There are no existing prosecutions to draw on in either jurisdiction, and future cases may rely on contextual evidence---where a model was circulated, how it was advertised, or accompanying prompts and filenames on a seized device. How to apply these laws where such context is weak or unavailable is an open question, and establishing capability is not the same as intent, as research suggests general-purpose models can be used to create explicit content \cite{wei2024exploring}. 

Determining which side of this line a tool falls on may therefore require measuring specialization to NCII and CSAM. Some routes to identifying specialization are addressed by the previous questions in this section, including training data (§\ref{subsubsec:data_detection}), weights or activations (§\ref{subsubsec:capability_detection}), and combined capabilities (§\ref{subsubsec:combination}). We discuss two more routes in this section.

The first is propensity---how readily a model produces such material relative to other content. Propensity measurement approaches in AI have often built on psychometrics, extending methods like item response theory (IRT) which infer latent traits from test responses. Previous work has extended IRT into a broader framework for measuring propensities \cite{romero2026capabilities}, and applied IRT to sandbagging detection and interpreting benchmark trends \cite{rivera2026item, spagliardi2026efficient}, but these works focus on language models rather than image generators.

The second is uplift---measuring how much a tool increases the effectiveness of producing NCII and CSAM beyond what is already possible. Previous work has proposed that legal liability for a finetuned model should track whether the finetune introduces a harmful capability that the base model does not have \cite{hacker2026regulation}. More broadly, \cite{kapoor2024position} introduce a framework to measure uplift of open weight models. However, they note that it is challenging to measure uplift (e.g. through randomized controlled trials) due to the criminality of this material. Tailoring uplift frameworks to these domains requires considering the impact on victims \cite{qiwei2026position}, which makes them distinct from other misuse vectors like fraud. Victim profiles also vary: \cite{cui2026celebrities} studied AI nudification dynamics on 4chan and found that non-celebrity individuals account for 55.8\% of targets, up from 4.7\% in prior studies, meaning nudification has shifted from public figures to individuals within a user’s social circle. 

Open questions include determining how much evidence of specialization is sufficient to establish intent, adapting psychometric methods to image generators, disentangling the contributions of base models and customization such as LoRAs, and refining methods so that evidence of `adaptation’ is reliable enough to be accepted by courts.

\subsection{Training Materials}
\label{subsec:question_training}

\begin{rqbox}
\subsubsection{How can the training materials for customizing models for NCII and CSAM be disrupted?}
\par
\begin{flushleft}
\functag{Enforcement}
\stacktag{Detection}
\end{flushleft}
\tcblower
\textbf{Scenario:} Law enforcement want to understand the ways that adversaries learn how to customize models to find out ways to disrupt and take down these training materials.
\end{rqbox}

Studies of adversarial communities show that the knowledge and capability to customize models for NCII and CSAM are shared through manuals and tutorials \cite{iwf2026aicsam, han2025characterizing}. These manuals are illegal under E\&W law. Identifying them would support enforcement, and disrupting the wider information-sharing ecosystem may reduce offending itself. Situational crime prevention suggests raising the effort required can reduce offending \cite{cornish2003opportunities}.

Information could be gathered and analyzed for disruption at two levels. The first is the ecosystem level. Research suggests adversaries concentrate around a few favored models and tools, so a systematic understanding of this landscape---and where know-how is exchanged---could offer valuable insight for operational disruption. Whether emerging techniques could be forecast early enough to support detection and takedown is unexplored.

The second is the artifact level. Where the tools adversaries use are known, the metadata those tools embed may support evidence-gathering. Training tools record metadata in released adapters, such as training configuration and dataset descriptors, and \cite{suriyakumar2026evaluation} found that the training hyperparameters of explicit LoRAs differed from those of non-explicit LoRAs. Whether such metadata signatures reliably indicate malicious use, or link an adapter to a particular guide or community, is an open question.
\subsection{Ethics and Research Governance}
\label{subsec:question_ethics}
 The technical problems discussed in §\ref{sec:technical} must be studied under the legal and ethical constraints described in §\ref{subsec:policylandscape}. No jurisdiction discussed so far offers researchers a widely operational and lawful path to studying whether systems are capable of generating CSAM.

While E\&W's Crime and Policing Act 2026 introduces a technology testing defence under Section 111, this defence has yet to be implemented.
Beyond legality, NCII and CSAM research carries risks not fully resolved by testing exemptions, including how to disclose findings without uplifting bad actors, draw valid conclusions from proxies, and protect NCII and CSAM researchers from exposure to harmful content. 
\begin{rqbox}
\subsubsection{How do we report on obfuscation research and update existing screening techniques without alerting adversaries?}
\par
\begin{flushleft}
\functag{Prevention}
\stacktag{Detection}
\end{flushleft}
\tcblower
\textbf{Scenario: }A research team has designed a new screening method to detect whether a model has been trained on malicious content. However, in their testing, they find that the method is vulnerable to certain types of attack. How does the research team share its method for other teams to implement and build upon, appropriately disclosing weaknesses without alerting perpetrators to effective evasion methods?
\end{rqbox}

Several problems above call for research into how NCII and CSAM producers avoid detection. Much of this research is inherently dual use given the same research that advances screening could also help to craft new evasion techniques. On one hand, research that documents current practices, such as stripping metadata, has marginal impact since they are already widely adopted. On the other hand, research that constructs novel attacks, such as backdoored LoRAs, may introduce new capabilities to would-be adversaries. Publishing a detector, or publishing the signal that a detector relies on, carries the risk that bad actors can act to evade it. 

Computer security has long dealt with these issues and manages these trade-offs through coordinated vulnerability disclosures: notifying the appropriate entity and delaying publication until a fix is available. This model does not translate well to NCII and CSAM for two reasons. First, there is no coordinating infrastructure comparable to a Computer Emergency Response Team (CERT/CC) for generative AI misuse, let alone for the domains discussed in this paper. Second, coordinated disclosure presupposes that those affected have the incentive and ability to rectify harms that have been identified. Many actors in the NCII and CSAM ecosystem, such as nudification services, instead profit from the capabilities researchers are seeking to identify, so notifying them invites evasion rather than remediation. 

Thus, open questions include who should be notified and how reports can be verified without generating harmful or illegal content, what mechanisms will lead to enforcement when there is no incentive to cooperate and if and when detectors should be released openly versus to vetted parties only. 

\begin{rqbox}
\subsubsection{How can we ensure research on proxies are transferable to conclusions about real threats?}
\par
\begin{flushleft}
\functag{Prevention}
\stacktag{Detection}
\end{flushleft}
\tcblower
\textbf{Scenario: }A research team has built a tool to detect CSAM-specialized models but cannot lawfully test it on real CSAM. They evaluate it on proxies and want to know whether strong results tell them anything reliable about real threats before passing it onto a team with sufficient authorization to test.
\end{rqbox}

The current practice is to develop mitigation tools and evaluate them on proxies, given that many categories NCII and CSAM content cannot be lawfully or ethically handled by researchers. Examples include adult nudity or child-related content in place of CSAM, and consenting adults in place of NCII victims. If this is to remain the state-of-affairs, research is needed to identify which proxies, if any, can be used to draw valid conclusions. There might be several reasons these proxies might not transfer: target content could be represented differently in model weights or bad actors could use different customization procedures than what proxies can simulate. This could lead researchers to overstate their results or miss important failure modes. Future work should therefore identify the conditions under which proxies will generalize to real NCII and CSAM, understand when they fail to generalize, and establish validation routes for authorized organizations to replicate proxy-based procedures on real data under Section 111.

\section{Discussion}

In this paper, we analyze existing NCII and CSAM laws in E\&W and the U.S. to identify areas of research that could help operationalize them. Our analysis shows that applying these laws to generative AI often depends on technical evidence or capabilities that current methods do not reliably provide. NCII and CSAM are distinct from other types of generative AI misuse in that the act of generating a new image is itself a revictimization, further complicating operationalization of laws \cite{qiwei2026position, iwf2026aicsam}. The resulting research agenda therefore focuses on the technical questions that arise in applying existing law. The questions range from algorithmic tests to help interpret legal definitions, to tools that prevent generation, to methods that track and disrupt harmful content.

Although our analysis uses these two jurisdictions as a starting point, many of the open questions generalize beyond them, as they focus on constructs common to other legal systems, such as identifiability and intent. We do not aim to provide an exhaustive overview of open questions, nor to propose legal reform. We also recognize the legal barriers to this research are real, and such research must be conducted responsibly. But these barriers are not a reason to avoid the work, which would let these harms grow unchecked. Taken together, these questions identify where technical research is needed for existing legal rules to be interpreted, implemented, and enforced in generative AI settings. In carrying this analysis out, we draw on adjacent fields such as privacy and propensity measurement while identify technical requirements that arise specifically from applying NCII and CSAM law to generative systems. 

\section*{Generative AI Usage Statement}

Generative AI was not used substantially for the content or analysis of this paper. It was used for the following tasks:
\begin{itemize}
    \item Literature search assistance: Claude Opus 5.5 was used to identify potentially relevant literature. All citations were manually reviewed and verified by the authors.
    \item Copyediting: Authors wrote all content. Claude Opus 4.8 was used to review grammar, correct typos, and improve the clarity of the content.
\end{itemize}

\section*{Ethics and Privacy Statement}

While the paper reflects on considerations for conducting NCII and CSAM research ethically, we acknowledge that documenting ambiguities in these laws could assist adversaries. To mitigate this, we consulted specialists in NCII and CSAM policy and law enforcement and shared a draft with them before submission so they could advise on sensitivities. We incorporated their feedback. We also only refer to publicly reported material when describing the tools and obfuscation techniques adversaries use. Furthermore, this work focuses on open questions at a conceptual level. We did not generate, access, or handle illegal material in producing the paper. Our overall assessment is that the benefits of disclosure in this paper outweigh the residual risks, as it is oriented towards enabling defensive research. To our knowledge, this is the first work to connect legal and policy analysis of NCII and CSAM with the technical research needed to operationalize it. 

\section*{Acknowledgements}
The authors would like to thank Amos Wittenberg, Andrew Strait, Cannelle Roumanie, Chris Summerfield, Punya Syon Pandey, and others for their feedback on this manuscript.

\bibliographystyle{ACM-Reference-Format}
\bibliography{bibliography}

@String{Computing = "Computing" }

@String{Computer = "{IEEE} Computer" }

@String{Springer = "Springer-Verlag" }

@misc{adobegenerativefill,
  author       = {{Adobe}},
  title        = {Generative Fill},
  howpublished = {\url{https://www.adobe.com/products/photoshop/generative-fill.html/}},
  note         = {Accessed: 22 September 2026},
  year={n.d.}
}

@misc{africa2026detecting,
  title={Detecting CSAM Text-to-Image LoRAs From Weights},
  author={Africa, David Demitri and Heine, Cate and Staes-Polet, Nadine and Mai, Kimberly},
  journal={arXiv preprint arXiv:2607.25750},
  year={2026}
}

@inproceedings{bertalmio2000image,
author = {Bertalmio, Marcelo and Sapiro, Guillermo and Caselles, Vincent and Ballester, Coloma},
title = {Image inpainting},
year = {2000},
isbn = {1581132085},
publisher = {ACM Press/Addison-Wesley Publishing Co.},
address = {USA},
url = {https://doi.org/10.1145/344779.344972},
doi = {10.1145/344779.344972},
booktitle = {Proceedings of the 27th Annual Conference on Computer Graphics and Interactive Techniques},
pages = {417–424},
numpages = {8},
series = {SIGGRAPH '00}
}

@inproceedings{bitouk2008face,
author = {Bitouk, Dmitri and Kumar, Neeraj and Dhillon, Samreen and Belhumeur, Peter and Nayar, Shree K.},
title = {Face swapping: automatically replacing faces in photographs},
year = {2008},
isbn = {9781450301121},
publisher = {Association for Computing Machinery},
address = {New York, NY, USA},
url = {https://doi.org/10.1145/1399504.1360638},
doi = {10.1145/1399504.1360638},
booktitle = {ACM SIGGRAPH 2008 Papers},
articleno = {39},
numpages = {8},
location = {Los Angeles, California},
series = {SIGGRAPH '08}
}

@inproceedings {carlini2023extracting,
author = {Nicolas Carlini and Jamie Hayes and Milad Nasr and Matthew Jagielski and Vikash Sehwag and Florian Tram{\`e}r and Borja Balle and Daphne Ippolito and Eric Wallace},
title = {Extracting Training Data from Diffusion Models},
booktitle = {32nd USENIX Security Symposium (USENIX Security 23)},
year = {2023},
isbn = {978-1-939133-37-3},
address = {Anaheim, CA},
pages = {5253--5270},
url = {https://www.usenix.org/conference/usenixsecurity23/presentation/carlini},
publisher = {USENIX Association},
month = aug
}

@inproceedings{chen2022trauma,
author = {Chen, Janet X. and McDonald, Allison and Zou, Yixin and Tseng, Emily and Roundy, Kevin A and Tamersoy, Acar and Schaub, Florian and Ristenpart, Thomas and Dell, Nicola},
title = {Trauma-Informed Computing: Towards Safer Technology Experiences for All},
year = {2022},
isbn = {9781450391573},
publisher = {Association for Computing Machinery},
address = {New York, NY, USA},
url = {https://doi.org/10.1145/3491102.3517475},
doi = {10.1145/3491102.3517475},
booktitle = {Proceedings of the 2022 CHI Conference on Human Factors in Computing Systems},
articleno = {544},
numpages = {20},
location = {New Orleans, LA, USA},
series = {CHI '22}
}

@article{cooper2025files,
  title={The files are in the computer: on copyright, memorization, and generative AI},
  author={Cooper, A Feder and Grimmelmann, James},
  journal={Chi.-Kent L. Rev.},
  volume={100},
  pages={141},
  year={2025},
  publisher={HeinOnline}
}

@article{cornish2003opportunities,
  title={Opportunities, precipitators and criminal decisions: A reply to Wortley's critique of situational crime prevention},
  author={Cornish, Derek Blaikie and Clarke, Ronald V},
  journal={Crime prevention studies},
  volume={16},
  pages={41--96},
  year={2003},
  publisher={Criminal Justice Press}
}

@misc{cuevas2026deepfake,
  title={Deepfake Pornography is Resilient to Regulatory and Platform Shocks},
  author={Cuevas, Alejandro and Ribeiro, Manoel Horta},
  journal={arXiv preprint arXiv:2602.02754},
  year={2026}
}

@misc{cui2026celebrities,
  title={From Celebrities to Anyone: Characterizing AI Nudification Content, Technology, and Community Dynamics on 4chan},
  author={Cui, Chi and Wu, Yixin and Zhang, Yang},
  journal={arXiv preprint arXiv:2606.27234},
  year={2026}
}

@misc{daffalla2026dual,
  title={Dual-Use AI Face Swap Apps Are Mostly Unsafe: A Systematic Safety Audit},
  author={Daffalla, Alaa and Chao, Sarah and Zeng, Eric},
  journal={arXiv preprint arXiv:2605.24735},
  year={2026}
}

@article{dan2023persistence,
  title={The persistence of knuckle creases during finger flexion for the identification of perpetrators from digital images of their hands},
  author={Dan, Lilly and Hackman, Lucina and Rinaldi, Vincenzo and Trucco, Emanuele},
  journal={Forensic Science International},
  volume={348},
  pages={111603},
  year={2023},
  publisher={Elsevier}
}

@InProceedings{deng2019arcface,
author = {Deng, Jiankang and Guo, Jia and Xue, Niannan and Zafeiriou, Stefanos},
title = {ArcFace: Additive Angular Margin Loss for Deep Face Recognition},
booktitle = {Proceedings of the IEEE/CVF Conference on Computer Vision and Pattern Recognition (CVPR)},
month = {June},
year = {2019},
publisher={IEEE},
address={Long Beach, CA, USA},
numpages={17}
}

@misc{deletang2024language,
  title={Language modeling is compression},
  author={Del{\'e}tang, Gr{\'e}goire and Ruoss, Anian and Duquenne, Paul-Ambroise and Catt, Elliot and Genewein, Tim and Mattern, Christopher and Grau-Moya, Jordi and Wenliang, Li Kevin and Aitchison, Matthew and Orseau, Laurent and others},
  booktitle={International Conference on Learning Representations},
  volume={2024},
  pages={14165--14181},
  year={2024}
}

@misc{enoughabuse,
  author       = {{ENOUGH ABUSE}},
  title        = {State Laws Criminalizing AI-Generated or Computer-Edited CSAM},
  howpublished = {\url{https://enoughabuse.org/get-vocal/laws-by-state/state-laws-criminalizing-ai-generated-or-computer-edited-child-sexual-abuse-material-csam//}},
  note         = {Accessed: 3 November 2026},
  year={2026}
}

@misc{
franceschelli2024training,
title={Training foundation models as data compression: On information, model weights and copyright law},
author={Franceschelli, Giorgio and Cevenini, Claudia and Musolesi, Mirco},
booktitle={The Second Workshop on Generative AI and Law},
year={2024},
url={https://blog.genlaw.org/pdfs/genlaw_icml2024/85.pdf}
}

@misc{
cooper2025extracting,
title={Extracting memorized pieces of (copyrighted) books from open-weight language models},
author={A. Feder Cooper and Mark Lemley and Allison Casasola and Ahmed M Ahmed and Aaron Gokaslan and Amy B. Cyphert and Christopher De Sa and Daniel E. Ho and Percy Liang},
booktitle={Third Conference on Language Modeling},
year={2026},
url={https://openreview.net/forum?id=VE3f9mjaCd}
}

@article{dravid2024interpreting,
  title={Interpreting the weight space of customized diffusion models},
  author={Dravid, Amil and Gandelsman, Yossi and Wang, Kuan-Chieh and Abdal, Rameen and Wetzstein, Gordon and Efros, Alexei A and Aberman, Kfir},
  journal={Advances in Neural Information Processing Systems},
  volume={37},
  pages={137334--137371},
  year={2024}
}

@misc{gan2025conceptmix++,
  title={ConceptMix++: Leveling the Playing Field in Text-to-Image Benchmarking via Iterative Prompt Optimization},
  author={Gan, Haosheng and Tinaz, Berk and Sepehri, Mohammad Shahab and Fabian, Zalan and Soltanolkotabi, Mahdi},
  journal={arXiv preprint arXiv:2507.03275},
  year={2025}
}

@inproceedings{gatys2016image,
  title={Image style transfer using convolutional neural networks},
  author={Gatys, Leon A and Ecker, Alexander S and Bethge, Matthias},
  booktitle={Proceedings of the IEEE Conference on Computer Vision and Pattern Recognition},
  pages={2414--2423},
  year={2016},
  publisher={IEEE},
  address={Las Vegas, NV, USA}
}

@inproceedings {gibson2025analyzing,
author = {Cassidy Gibson and Daniel Olszewski and Natalie Grace Brigham and Anna Crowder and Kevin R. B. Butler and Patrick Traynor and Elissa M. Redmiles and Tadayoshi Kohno},
title = {Analyzing the {AI} Nudification Application Ecosystem},
booktitle = {34th USENIX Security Symposium (USENIX Security 25)},
year = {2025},
isbn = {978-1-939133-52-6},
address = {Seattle, WA},
pages = {1--20},
url = {https://www.usenix.org/conference/usenixsecurity25/presentation/gibson},
publisher = {USENIX Association},
month = aug
}

@inproceedings{gubri2024trap,
  title={Trap: Targeted random adversarial prompt honeypot for black-box identification},
  author={Gubri, Martin and Ulmer, Dennis and Lee, Hwaran and Yun, Sangdoo and Oh, Seong Joon},
  booktitle={Findings of the Association for Computational Linguistics: ACL 2024},
  pages={11496--11517},
  year={2024},
  publisher={Association for Computational Linguistics},
  address={Bangkok, Thailand}
}

@inproceedings{jin2024proflingo,
  title={Proflingo: A fingerprinting-based intellectual property protection scheme for large language models},
  author={Jin, Heng and Zhang, Chaoyu and Shi, Shanghao and Lou, Wenjing and Hou, Y Thomas},
  booktitle={2024 IEEE Conference on Communications and Network Security (CNS)},
  pages={1--9},
  year={2024},
  organization={IEEE},
  publisher={IEEE},
  address={Taipei, Taiwan}
}

@inproceedings {han2025characterizing,
author = {Catherine Han and Anne Li and Deepak Kumar and Zakir Durumeric},
title = {Characterizing the {MrDeepFakes} Sexual Deepfake Marketplace},
booktitle = {34th USENIX Security Symposium (USENIX Security 25)},
year = {2025},
isbn = {978-1-939133-52-6},
address = {Seattle, WA},
pages = {5169--5188},
url = {https://www.usenix.org/conference/usenixsecurity25/presentation/han},
publisher = {USENIX Association},
month = aug
}

@misc{zhang2024promptla,
  title={PromptLA: Towards Integrity Verification of Black-box Text-to-Image Diffusion Models},
  author={Zhang, Zhuomeng and Li, Fangqi and Di, Chong and Zhu, Hongyu and Wang, Hanyi and Wang, Shilin},
  journal={arXiv preprint arXiv:2412.16257},
  year={2024}
}

@inproceedings{yang2026challenge,
    title = "The Challenge of Identifying the Origin of Black-Box Large Language Models",
    author = "Yang, Ziqing  and
      Wu, Yixin  and
      Shen, Yun  and
      Dai, Wei  and
      Backes, Michael  and
      Zhang, Yang",
    editor = "Habernal, Ivan  and
      Ghanavati, Sepideh  and
      Haghighi, Sara  and
      Ramesh, Krithika  and
      Igamberdiev, Timour  and
      Wilson, Shomir",
    booktitle = "Proceedings of the Seventh Workshop on Privacy in Natural Language Processing",
    month = jul,
    year = "2026",
    address = "San Diego, California",
    publisher = "Association for Computational Linguistics",
    url = "https://aclanthology.org/2026.privatenlp-main.2/",
    doi = "10.18653/v1/2026.privatenlp-main.2",
    pages = "7--25",
    ISBN = "979-8-89176-397-5",
}

@misc{guo2024one,
  title={One prompt to verify your models: Black-box text-to-image models verification via non-transferable adversarial attacks},
  author={Guo, Ji and Jiang, Wenbo and Zhang, Rui and Lu, Guoming and Li, Hongwei},
  journal={arXiv preprint arXiv:2410.22725},
  year={2024}
}

@misc{googlemagiceraser,
  author       = {{Google}},
  title        = {Magic Eraser},
  howpublished = {\url{https://www.google.com/intl/en_us/photos/editing/}},
  note         = {Accessed: 22 September 2026},
  year={n.d.},
}

@article{gorwa2024moderating,
  title={Moderating model marketplaces: Platform governance puzzles for AI intermediaries},
  author={Gorwa, Robert and Veale, Michael},
  journal={Law, Innovation and Technology},
  volume={16},
  number={2},
  pages={341--391},
  year={2024},
  publisher={Taylor \& Francis}
}

@inproceedings{cao2021ipguard,
author = {Cao, Xiaoyu and Jia, Jinyuan and Gong, Neil Zhenqiang},
title = {IPGuard: Protecting Intellectual Property of Deep Neural Networks via Fingerprinting the Classification Boundary},
year = {2021},
isbn = {9781450382878},
publisher = {Association for Computing Machinery},
address = {New York, NY, USA},
url = {https://doi.org/10.1145/3433210.3437526},
doi = {10.1145/3433210.3437526},
booktitle = {Proceedings of the 2021 ACM Asia Conference on Computer and Communications Security},
pages = {14–25},
numpages = {12},
location = {Virtual Event, Hong Kong},
series = {ASIA CCS '21}
}

@INPROCEEDINGS{chen2022copy,
  author={Chen, Jialuo and Wang, Jingyi and Peng, Tinglan and Sun, Youcheng and Cheng, Peng and Ji, Shouling and Ma, Xingjun and Li, Bo and Song, Dawn},
  booktitle={2022 IEEE Symposium on Security and Privacy (SP)}, 
  title={Copy, Right? A Testing Framework for Copyright Protection of Deep Learning Models}, 
  year={2022},
  pages={824-841},
  address   = {San Francisco, CA, USA},
  publisher = {IEEE},
  doi={10.1109/SP46214.2022.9833747}}

@inproceedings{jia2021zest,
  title={A zest of lime: Towards architecture-independent model distances},
  author={Jia, Hengrui and Chen, Hongyu and Guan, Jonas and Shamsabadi, Ali Shahin and Papernot, Nicolas},
  booktitle={International Conference on Learning Representations},
  year={2021},
  publisher={OpenReview},
  address={Virtual},
  numpages={17}
}

@inproceedings{chen2022teacher,
author = {Yufei Chen and Chao Shen and Cong Wang and Yang Zhang},
title = {Teacher Model Fingerprinting Attacks Against Transfer Learning},
booktitle = {31st USENIX Security Symposium (USENIX Security 22)},
year = {2022},
isbn = {978-1-939133-31-1},
address = {Boston, MA},
pages = {3593--3610},
url = {https://www.usenix.org/conference/usenixsecurity22/presentation/chen-yufei},
publisher = {USENIX Association},
month = aug
}

@misc{ding2026stop,
  title={How to stop playing whack-a-mole: Mapping the ecosystem of technologies facilitating ai-generated non-consensual intimate images},
  author={Ding, Michelle L and Suresh, Harini and Venkatasubramanian, Suresh},
  journal={arXiv preprint arXiv:2602.04759},
  year={2026}
}

@InProceedings{duan2023diffusion,
  title = 	 {Are Diffusion Models Vulnerable to Membership Inference Attacks?},
  author =       {Duan, Jinhao and Kong, Fei and Wang, Shiqi and Shi, Xiaoshuang and Xu, Kaidi},
  booktitle = 	 {Proceedings of the 40th International Conference on Machine Learning},
  pages = 	 {8717--8730},
  year = 	 {2023},
  editor = 	 {Krause, Andreas and Brunskill, Emma and Cho, Kyunghyun and Engelhardt, Barbara and Sabato, Sivan and Scarlett, Jonathan},
  volume = 	 {202},
  series = 	 {Proceedings of Machine Learning Research},
  month = 	 {23--29 Jul},
  publisher =    {PMLR},
  url = 	 {https://proceedings.mlr.press/v202/duan23b.html},
  address = {Honolulu, Hawaii, USA},
}

@misc{dubinski2024towards,
  title={Towards more realistic membership inference attacks on large diffusion models},
  author={Dubi{\'n}ski, Jan and Kowalczuk, Antoni and Pawlak, Stanis{\l}aw and Rokita, Przemyslaw and Trzci{\'n}ski, Tomasz and Morawiecki, Pawe{\l}},
  booktitle={Proceedings of the IEEE/CVF Winter Conference on Applications of Computer Vision},
  pages={4860--4869},
  year={2024}
}

@article{gao2025toward,
  title={Toward dataset copyright evasion attack against personalized text-to-image diffusion models},
  author={Gao, Kuofeng and Zhu, Yufei and Li, Yiming and Bai, Jiawang and Yang, Yong and Li, Zhifeng and Xia, Shu-Tao},
  journal={IEEE Transactions on Information Forensics and Security},
  volume={21},
  pages={725--740},
  year={2025},
  publisher={IEEE}
}

@inproceedings{ha2026identifying,
author = {Anna Yoo Jeong Ha and Wenxin Ding and Stanley Wu and Shawn Shan and Haitao Zheng and Ben Y. Zhao},
title = {Identifying Provenance of Generative {Text-to-Image} Models},
booktitle = {35th USENIX Security Symposium (USENIX Security 26)},
year = {2026},
isbn = {978-1-939133-58-8},
address = {Baltimore, MD},
pages = {4503--4522},
url = {https://www.usenix.org/conference/usenixsecurity26/presentation/ha},
publisher = {USENIX Association},
month = aug
}

@article{hacker2026regulation,
  title={The regulation of fine-tuning: Federated compliance for modified general-purpose AI models},
  author={Hacker, Philipp and Holweg, Matthias},
  journal={Computer Law \& Security Review},
  volume={60},
  pages={106234},
  year={2026},
  publisher={Elsevier}
}

@misc{hamburg2024llms,
  author       = {{The Hamburg Commissioner for Data Protection and Freedom of Information}},
  title        = {Discussion Paper: Large Language Models and Personal Data},
  howpublished = {\url{https://datenschutz-hamburg.de/fileadmin/user_upload/HmbBfDI/Datenschutz/Informationen/240715_Discussion_Paper_Hamburg_DPA_KI_Models.pdf/}},
  year         = {2024},
  note         = {Accessed: 28 September 2026},
  organization = {The Hamburg Commissioner for Data Proteciton and Freedom of Information}
}

@inproceedings{hawkins2025deepfakes,
author = {Hawkins, Will and Mittelstadt, Brent and Russell, Chris},
title = {Deepfakes on Demand: The rise of accessible non-consensual deepfake image generators},
year = {2025},
isbn = {9798400714825},
publisher = {Association for Computing Machinery},
address = {New York, NY, USA},
url = {https://doi.org/10.1145/3715275.3732107},
doi = {10.1145/3715275.3732107},
booktitle = {Proceedings of the 2025 ACM Conference on Fairness, Accountability, and Transparency},
pages = {1602–1614},
numpages = {13},
location = {Athens, Greece},
series = {FAccT '25}
}

@misc{
hertz2023prompttoprompt,
title={Prompt-to-Prompt Image Editing with Cross-Attention Control},
author={Amir Hertz and Ron Mokady and Jay Tenenbaum and Kfir Aberman and Yael Pritch and Daniel Cohen-Or},
booktitle={The Eleventh International Conference on Learning Representations },
year={2023},
url={https://openreview.net/forum?id=_CDixzkzeyb}
}

@inproceedings{hrkac2016tattoo,
  title={Tattoo detection for soft biometric de-identification based on convolutional neural networks},
  author={Hrkac, Tomislav and Brkic, Karla and Kalafatic, Zoran},
  booktitle={Proc. OAGM-ARW Joint Workshop},
  publisher={OAGM},
  address={Austria},
  pages={131--138},
  year={2016}
}

@article{ho2022video,
  title={Video diffusion models},
  author={Ho, Jonathan and Salimans, Tim and Gritsenko, Alexey and Chan, William and Norouzi, Mohammad and Fleet, David J},
  journal={Advances in neural information processing systems},
  volume={35},
  pages={8633--8646},
  year={2022}
}

@inproceedings{
hu2022lora,
title={Lo{RA}: Low-Rank Adaptation of Large Language Models},
author={Edward J Hu and yelong shen and Phillip Wallis and Zeyuan Allen-Zhu and Yuanzhi Li and Shean Wang and Lu Wang and Weizhu Chen},
booktitle={International Conference on Learning Representations},
year={2022},
url={https://openreview.net/forum?id=nZeVKeeFYf9},
publisher={OpenReview},
address={Virtual},
numpages={13}
}

@misc{hu2024stable,
  title={Stable signature is unstable: Removing image watermark from diffusion models},
  author={Hu, Yuepeng and Jiang, Zhengyuan and Guo, Moyang and Gong, Neil},
  journal={arXiv preprint arXiv:2405.07145},
  year={2024}
}

@techreport{icmec2023csam,
  author      = {{International Centre for Missing \& Exploited Children (ICMEC)}},
  title       = {Child Sexual Abuse Material: Model Legislation \& Global Review},
  institution = {International Centre for Missing \& Exploited Children},
  edition     = {9th},
  year        = {2023},
  note        = {Accessed: 22 September 2026}
}

@article{parti2024legal,
  title={The legal challenges of realistic and AI-driven child sexual abuse material: regulatory and enforcement perspectives in Europe},
  author={Parti, Katalin and Szab{\'o}, Judit},
  journal={Laws},
  volume={13},
  number={6},
  pages={67},
  year={2024},
  publisher={MDPI}
}

@misc{isd2026mapping,
  author       = {{Institute for Strategic Dialogue}},
  title        = {Mapping the `Nudify' Tools Ecosystem},
  howpublished = {\url{https://www.isdglobal.org/publication/mapping-the-nudify-tools-ecosystem/}},
  year         = {2026},
  note         = {Accessed: 22 September 2026},
  organization = {Institute for Strategic Dialogue}
}

@misc{isd2026ncii,
  author       = {{Institute for Strategic Dialogue}},
  title        = {Non-Consensual Intimate Imagery},
  howpublished = {\url{https://www.isdglobal.org/isd-explainer/non-consensual-intimate-imagery-ncii-abuse/}},
  year         = {2026},
  note         = {Accessed: 2 November 2026},
  organization = {Institute for Strategic Dialogue}
}

@misc{iwf2026aicsam,
  author       = {{Internet Watch Foundation}},
  title        = {AI-Generated Child Sexual Abuse 2026: 2026 Report on Trends, Data \& Human Impact},
  howpublished = {\url{https://www.iwf.org.uk/about-us/why-we-exist/our-research/how-ai-is-being-abused-to-create-child-sexual-abuse-imagery}},
  year         = {2026},
  note         = {Accessed: 22 September 2026},
  organization = {Internet Watch Foundation}
}

@misc{doj2026citizensguide,
  author       = {Criminal Division, U.S. Department of Justice},
  title        = {Citizen's Guide To U.S. Federal Child Exploitation And Obscenity Laws},
  howpublished = {\url{https://www.justice.gov/criminal/criminal-ceos/citizens-guide-us-federal-child-exploitation-and-obscenity-laws}},
  year         = {2026},
  note         = {Accessed: 28 September 2026},
  organization = {U.S. Department of Justice}
}

@inproceedings{
kapoor2024position,
title={Position: On the Societal Impact of Open Foundation Models},
author={Sayash Kapoor and Rishi Bommasani and Kevin Klyman and Shayne Longpre and Ashwin Ramaswami and Peter Cihon and Aspen K Hopkins and Kevin Bankston and Stella Biderman and Miranda Bogen and Rumman Chowdhury and Alex Engler and Peter Henderson and Yacine Jernite and Seth Lazar and Stefano Maffulli and Alondra Nelson and Joelle Pineau and Aviya Skowron and Dawn Song and Victor Storchan and Daniel Zhang and Daniel E. Ho and Percy Liang and Arvind Narayanan},
booktitle={International Conference on Machine Learning},
year={2024},
url={https://openreview.net/forum?id=jRX6yCxFhx},
publisher={OpenReview},
address={Vienna, Austria},
numpages={23}
}

@inproceedings{
kwon2023diffusion,
title={Diffusion Models Already Have A Semantic Latent Space},
author={Mingi Kwon and Jaeseok Jeong and Youngjung Uh},
booktitle={The Eleventh International Conference on Learning Representations},
year={2023},
url={https://openreview.net/forum?id=pd1P2eUBVfq},
publisher={OpenReview},
address={Kigali, Rwanda},
numpages={35}
}

@conference{openproblemschildsafety2026,
  title={Preventing AI-Generated Child Sexual Abuse Material Necessitates New Approaches to AI Safety},
  author={Kale, Neil and Portnoff, Rebecca and Thaker, Pratiksha and Simpson, Michael and Wang, Rob and Kuo, Kevin and Yadav, Chhavi and Smith, Virginia},
  booktitle={International Conference on Machine Learning},
  year={2026},
  publisher={ICLR},
  address={Rio de Janeiro, Brazil},
  numpages={21}
}

@misc{kamachee2025video,
  title={Video deepfake abuse: How company choices predictably shape misuse patterns},
  author={Kamachee, Max and Casper, Stephen and Ding, Michelle L and Yew, Rui-Jie and Reuel, Anka and Biderman, Stella and Hadfield-Menell, Dylan},
  journal={arXiv preprint arXiv:2512.11815},
  year={2025}
}

@inproceedings{kang2026if,
author = {Kang, Chaewon and Lee, Youjin and Kwak, Haewoon and An, Jisun and Han, Jinyoung},
title = {If I Change Your Eyes, Are You Still You? Identity Thresholds in Deepfake Perception},
year = {2026},
isbn = {9798400722813},
publisher = {Association for Computing Machinery},
address = {New York, NY, USA},
url = {https://doi.org/10.1145/3772363.3798341},
doi = {10.1145/3772363.3798341},
booktitle = {Proceedings of the Extended Abstracts of the 2026 CHI Conference on Human Factors in Computing Systems},
articleno = {390},
numpages = {5},
location = {Barcelona, Spain},
series = {CHI EA '26}
}

@inproceedings{kazemi2014one,
  title={One millisecond face alignment with an ensemble of regression trees},
  author={Kazemi, Vahid and Sullivan, Josephine},
  booktitle={Proceedings of the IEEE Conference on Computer Vision and Pattern Recognition (CVPR)},
  pages={1867--1874},
  year={2014},
  publisher={IEEE},
  address={Columbus, Ohio}
}

@inproceedings{li2017generative,
  title={Generative face completion},
  author={Li, Yijun and Liu, Sifei and Yang, Jimei and Yang, Ming-Hsuan},
  booktitle={2017 IEEE Conference on Computer Vision and Pattern Recognition (CVPR)},
  pages={5892--5900},
  year={2017},
  organization={IEEE},
  publisher={IEEE},
  address={Honolulu, Hawaii}
}

@misc{li2019faceshifter,
  title={Faceshifter: Towards high fidelity and occlusion aware face swapping},
  author={Li, Lingzhi and Bao, Jianmin and Yang, Hao and Chen, Dong and Wen, Fang},
  journal={arXiv preprint arXiv:1912.13457},
  year={2019}
}

@inproceedings{li2024photomaker,
  title={Photomaker: Customizing realistic human photos via stacked id embedding},
  author={Li, Zhen and Cao, Mingdeng and Wang, Xintao and Qi, Zhongang and Cheng, Ming-Ming and Shan, Ying},
  booktitle={2024 IEEE/CVF Conference on Computer Vision and Pattern Recognition (CVPR)},
  pages={8640--8650},
  year={2024},
  organization={IEEE},
  publisher={IEEE},
  address={Seattle, Washington, USA}
}

@InProceedings{longpre2024position,
  title = 	 {Position: A Safe Harbor for {AI} Evaluation and Red Teaming},
  author =       {Longpre, Shayne and Kapoor, Sayash and Klyman, Kevin and Ramaswami, Ashwin and Bommasani, Rishi and Blili-Hamelin, Borhane and Huang, Yangsibo and Skowron, Aviya and Yong, Zheng Xin and Kotha, Suhas and Zeng, Yi and Shi, Weiyan and Yang, Xianjun and Southen, Reid and Robey, Alexander and Chao, Patrick and Yang, Diyi and Jia, Ruoxi and Kang, Daniel and Pentland, Alex and Narayanan, Arvind and Liang, Percy and Henderson, Peter},
  booktitle = 	 {Proceedings of the 41st International Conference on Machine Learning},
  pages = 	 {32691--32710},
  year = 	 {2024},
  editor = 	 {Salakhutdinov, Ruslan and Kolter, Zico and Heller, Katherine and Weller, Adrian and Oliver, Nuria and Scarlett, Jonathan and Berkenkamp, Felix},
  volume = 	 {235},
  series = 	 {Proceedings of Machine Learning Research},
  month = 	 {21--27 Jul},
  publisher =    {PMLR},
  url = 	 {https://proceedings.mlr.press/v235/longpre24a.html},
  address={Vienna, Austria}
  }

@article{lu2026concepts,
  title={When Are Concepts Erased From Diffusion Models?},
  author={Lu, Kevin and Kriplani, Nicky and Gandikota, Rohit and Pham, Minh and Bau, David and Hegde, Chinmay and Cohen, Niv},
  journal={Advances in Neural Information Processing Systems},
  volume={38},
  pages={14751--14775},
  year={2026}
}

@inproceedings{luo2025privacy,
  title={Privacy-preserving low-rank adaptation against membership inference attacks for latent diffusion models},
  author={Luo, Zihao and Xu, Xilie and Liu, Feng and Koh, Yun Sing and Wang, Di and Zhang, Jingfeng},
  booktitle={Proceedings of the AAAI Conference on Artificial Intelligence},
  volume={39},
  pages={5883--5891},
  year={2025},
  publisher={AAAI},
  address={Philadelpha, Pennsylvania, USA}
}

@misc{ma2026safegen,
  title={SafeGen-Bench: Benchmarking Safety in Image-Conditioned Text-to-Video Generation},
  author={Ma, Yingzi and Liu, Xiaogeng and Zheng, Yawen and Xiao, Chaowei},
  journal={arXiv preprint arXiv:2606.01481},
  year={2026}
}

@article{naruniec2020high,
  title={High-resolution neural face swapping for visual effects},
  author={Naruniec, Jacek and Helminger, Leonhard and Schroers, Christopher and Weber, Romann M},
  journal={Computer Graphics Forum},
  volume={39},
  number={4},
  pages={173--184},
  year={2020},
  organization={Wiley Online Library}
}

@article{nurhudatiana2013individuality,
  title={The individuality of relatively permanent pigmented or vascular skin marks (RPPVSM) in independently and uniformly distributed patterns},
  author={Nurhudatiana, Arfika and Kong, Adams Wai-Kin and Matinpour, Keyan and Chon, Deborah and Altieri, Lisa and Cho, Siu-Yeung and Craft, Noah},
  journal={IEEE Transactions on Information Forensics and Security},
  volume={8},
  number={6},
  pages={998--1012},
  year={2013},
  publisher={IEEE}
}

@misc{ofcom2026guidance,
  author       = {{Ofcom}},
  title        = {Illegal content duties under the Online Safety Act},
  howpublished = {\url{https://www.ofcom.org.uk/online-safety/illegal-and-harmful-content/illegal-content-duties-under-the-online-safety-act/}},
  year         = {2026},
  note         = {Accessed: 22 September 2026},
  organization = {Ofcom}
}

@article{okawa2023compositional,
  title={Compositional abilities emerge multiplicatively: Exploring diffusion models on a synthetic task},
  author={Okawa, Maya and Lubana, Ekdeep S and Dick, Robert and Tanaka, Hidenori},
  journal={Advances in Neural Information Processing Systems},
  volume={36},
  pages={50173--50195},
  year={2023}
}

@inproceedings{pathak2016context,
  title={Context encoders: Feature learning by inpainting},
  author={Pathak, Deepak and Krahenbuhl, Philipp and Donahue, Jeff and Darrell, Trevor and Efros, Alexei A},
  booktitle={Proceedings of the IEEE conference on computer vision and pattern recognition},
  pages={2536--2544},
  year={2016},
  publisher={IEEE},
  address={Las Vegas, Nevada, USA}
}

@inproceedings{
qiwei2026position,
title={Position: {AI}/{ML} Deepfake Research is Misaligned with {AI} Generated Non-Consensual Intimate Imagery ({AIG}-{NCII})},
author={Li Qiwei and Wells Lucas Santo and Sarita Schoenebeck and Eric Gilbert},
booktitle={Forty-third International Conference on Machine Learning Position Paper Track},
year={2026},
url={https://openreview.net/forum?id=mLhZzo7BIb},
publisher={OpenReview},
address={Seoul, South Korea},
numpages={15}
}

@misc{qi2026cert,
  title={Cert-LAS: Toward Certified Model Ownership Verification for Text-to-Image Diffusion Models via Layer-Adaptive Smoothing},
  author={Qi, Leyi and Li, Yiming and Liang, Siyuan and Tu, Zhengzhong and Tao, Dacheng},
  journal={arXiv preprint arXiv:2605.29809},
  year={2026}
}

@misc{qiu2026discovering,
  title={Discovering Undisclosed Concepts in LoRA Adapters for Diffusion-Based AIGC Models},
  author={Qiu, Yupeng and Fang, Han and Sharma, Dhananjai and Chang, Ee-Chien},
  journal={arXiv preprint arXiv:2610.18990},
  year={2026}
}

@misc{ran2025lora,
  title={Lora-leak: Membership inference attacks against lora fine-tuned language models},
  author={Ran, Delong and He, Xinlei and Cong, Tianshuo and Wang, Anyu and Li, Qi and Wang, Xiaoyun},
  journal={arXiv preprint arXiv:2507.18302},
  year={2025}
}

@misc{rivera2026item,
  title={Item Response Theory for AI Safety},
  author={Rivera, Joshua Fonseca and Shah, Neil and Africa, David Demitri and Voudouris, Konstantinos},
  journal={arXiv preprint arXiv:2608.05086},
  year={2026}
}

@inproceedings{rombach2022high,
  title={High-resolution image synthesis with latent diffusion models},
  author={Rombach, Robin and Blattmann, Andreas and Lorenz, Dominik and Esser, Patrick and Ommer, Bj{\"o}rn},
  booktitle={2022 IEEE/CVF conference on computer vision and pattern recognition (CVPR)},
  pages={10674--10685},
  year={2022},
  organization={IEEE},
  publisher={IEEE},
  address={New Orleans, Louisiana, USA}
}

@misc{romero2026capabilities,
  title={Capabilities Ain't All You Need: Measuring Propensities in AI},
  author={Romero-Alvarado, Daniel and Mart{\'\i}nez-Plumed, Fernando and Pacchiardi, Lorenzo and Save, Hugo and Pawar, Siddhesh Milind and Mehrbakhsh, Behzad and Casares, Pablo Antonio Moreno and Slater, Ben and Bova, Paolo and Romero, Peter and others},
  journal={arXiv preprint arXiv:2602.18182},
  year={2026}
}

@inproceedings{
shahreza2024unveiling,
title={Unveiling Synthetic Faces: How Synthetic Datasets Can Expose Real Identities},
author={Hatef Otroshi Shahreza and S{\'e}bastien Marcel},
booktitle={The Third Workshop on New Frontiers in Adversarial Machine Learning},
year={2024},
url={https://openreview.net/forum?id=oWDlWdxrM8},
publisher={OpenReview},
address={Vancouver, Canada},
numpages={12}
}

@misc{spagliardi2026efficient,
  title={Efficient Safety Benchmarking via Item Response Theory},
  author={Spagliardi, Fabio and Silva, M{\'\i}rian and Datta, Ayan and Zhou, Aiden and Bonagiri, Vamshi and Cruz, Diogo},
  journal={arXiv preprint arXiv:2606.20626},
  year={2026}
}

@misc{stopncii,
  author       = {{StopNCII.org}},
  title        = {StopNCII.org},
  howpublished = {\url{https://stopncii.org/}},
  year         = {2026},
  note         = {Accessed: 27 October 2026},
  organization = {StopNCII}
}

@misc{suriyakumar2026evaluation,
  title={Evaluation without Generation: Non-Generative Assessment of Harmful Model Specialization with Applications to CSAM},
  author={Suriyakumar, Vinith M and Sekhari, Ayush and Stempfle, Lena and Wang, Robertson and Simpson, Michael and Portnoff, Rebecca and Ghassemi, Marzyeh and Wilson, Ashia C},
  journal={arXiv preprint arXiv:2604.25119},
  year={2026}
}

@misc{wang2024instantid,
  title={Instantid: Zero-shot identity-preserving generation in seconds},
  author={Wang, Qixun and Bai, Xu and Wang, Haofan and Qin, Zekui and Chen, Anthony and Li, Huaxia and Tang, Xu and Hu, Yao},
  journal={arXiv preprint arXiv:2401.07519},
  year={2024}
}

@inproceedings{wei2024exploring,
  title={Exploring the use of abusive generative AI models on Civitai},
  author={Wei, Yiluo and Zhu, Yiming and Hui, Pan and Tyson, Gareth},
  booktitle={Proceedings of the 32nd ACM International Conference on Multimedia},
  pages={6949--6958},
  year={2024},
  publisher={Association for Computing Machinery},
  address={Melbourne, Australia}
}

@inproceedings{tang2023daam,
  title={What the daam: Interpreting stable diffusion using cross attention},
  author={Tang, Raphael and Liu, Linqing and Pandey, Akshat and Jiang, Zhiying and Yang, Gefei and Kumar, Karun and Stenetorp, Pontus and Lin, Jimmy and T{\"u}re, Ferhan},
  booktitle={Proceedings of the 61st Annual Meeting of the Association for Computational Linguistics (Volume 1: Long Papers)},
  pages={5644--5659},
  year={2023},
  publisher={Association for Computational Linguistics},
  address={Toronto, Canada}
}

@inproceedings{vora2025identity,
  title={Identity-focused inference and extraction attacks on diffusion models},
  author={Vora, Jayneel and Bouacida, Nader and Krishnan, Aditya and Shankar, Prabhu and Mohapatra, Prasant},
  booktitle={Proceedings of the 40th ACM/SIGAPP Symposium on Applied Computing},
  pages={1522--1530},
  year={2025},
  publisher={Association for Computing Machinery},
  address={New York, United States}
}

@article{wu2024conceptmix,
  title={Conceptmix: A compositional image generation benchmark with controllable difficulty},
  author={Wu, Xindi and Yu, Dingli and Huang, Yangsibo and Russakovsky, Olga and Arora, Sanjeev},
  journal={Advances in Neural Information Processing Systems},
  volume={37},
  pages={86004--86047},
  year={2024}
}

@misc{wu2024leveraging,
  title={Leveraging model guidance to extract training data from personalized diffusion models},
  author={Wu, Xiaoyu and Zhang, Jiaru and Wu, Zhiwei Steven},
  journal={arXiv preprint arXiv:2410.03039},
  year={2024}
}

@inproceedings{ruder2016artistic,
  title={Artistic style transfer for videos},
  author={Ruder, Manuel and Dosovitskiy, Alexey and Brox, Thomas},
  booktitle={German Conference on Pattern Recognition},
  pages={26--36},
  year={2016},
  organization={Springer},
  publisher={Springer},
  address={Hannover, Germany}
}

@inproceedings{wang2025sleepermark,
  title={Sleepermark: Towards robust watermark against fine-tuning text-to-image diffusion models},
  author={Wang, Zilan and Guo, Junfeng and Zhu, Jiacheng and Li, Yiming and Huang, Heng and Chen, Muhao and Tu, Zhengzhong},
  booktitle={2025 IEEE/CVF Conference on Computer Vision and Pattern Recognition (CVPR)},
  pages={8213--8224},
  year={2025},
  organization={IEEE},
  publisher={IEEE},
  address={Nashville, Tennessee, USA}
}

@misc{yao2024risks,
  title={Risks when sharing lora fine-tuned diffusion model weights},
  author={Yao, Dixi},
  journal={arXiv preprint arXiv:2409.08482},
  year={2024}
}

@inproceedings{yuan2025lurks,
  title={What lurks within? concept auditing for shared diffusion models at scale},
  author={Yuan, Xiaoyong and Ma, Xiaolong and Guo, Linke and Zhang, Lan},
  booktitle={Proceedings of the 2025 ACM SIGSAC Conference on Computer and Communications Security},
  pages={4154--4168},
  year={2025},
  publisher={Association for Computing Machinery},
  address={Taipei, Taiwan}
}

@inproceedings{zhang2024avid,
  title={Avid: Any-length video inpainting with diffusion model},
  author={Zhang, Zhixing and Wu, Bichen and Wang, Xiaoyan and Luo, Yaqiao and Zhang, Luxin and Zhao, Yinan and Vajda, Peter and Metaxas, Dimitris and Yu, Licheng},
  booktitle={2024 IEEE/CVF Conference on Computer Vision and Pattern Recognition (CVPR)},
  pages={7162--7172},
  year={2024},
  organization={IEEE},
  publisher={IEEE},
  address={Seattle, Washington, USA}
}

@inproceedings{zhou2023propainter,
  title={Propainter: Improving propagation and transformer for video inpainting},
  author={Zhou, Shangchen and Li, Chongyi and Chan, Kelvin CK and Loy, Chen Change},
  booktitle={Proceedings of the IEEE/CVF international conference on computer vision},
  pages={10477--10486},
  year={2023},
  publisher={IEEE},
  address={Paris, France}
}

\appendix

\section{Generation Techniques}
\label{appendix:generation_techniques}
\begin{itemize}
    \item \textbf{Face swap}. Face swaps transfer a target individual’s face onto a source image or video. Early methods located faces using facial landmarks \cite{kazemi2014one} and replaced them using geometric warping \cite{bitouk2008face}. More recent methods learn a representation of the target’s face, which preserves their identity while adapting to the lighting of the source and maintaining temporal consistency across video frames \cite{naruniec2020high, li2019faceshifter}. For NCII and CSAM, adversaries use face swaps to superimpose a victim’s face onto existing nude imagery \cite{isd2026mapping}. An audit of face swap services found that 70\% had no technical safeguards against the generation of nude images \cite{daffalla2026dual}.
    \item \textbf{Inpainting}. Inpainting replaces a selected region of an image or video with newly generated content \cite{bertalmio2000image, zhou2023propainter}. This is the visual equivalent of `filling in the blanks’ in a sentence. An inpainting model can specialize in replacing the gaps with a specific type of content or can be steered with a text prompt \cite{pathak2016context, li2017generative, zhang2024avid}. Inpainting has legitimate uses and is now standard in photo editing services \cite{googlemagiceraser, adobegenerativefill}, but the same capability could be used for clothes removal \cite{gibson2025analyzing}. 
    \item \textbf{Style transfer}. Style transfer changes an input image or video into a different visual style, for example, mimicking a particular artist \cite{gatys2016image, ruder2016artistic} or filters. Unlike inpainting, which changes a selected region, style transfer transforms the whole image. This can be used to render an existing image in a nude or sexualized style \cite{gibson2025analyzing}.
    \item \textbf{Complete synthetic generation}. These generators create images or video entirely from scratch and are typically steered using a text prompt \cite{rombach2022high, ho2022video}. Fully synthetic content may not include a real person at all. Diffusion-type models are often the dominant choice for generating NCII and CSAM, with one study finding that open-weight video generation models are disproportionately used to create explicit content \cite{kamachee2025video}.

\end{itemize}

\end{document}